\newif\ifsubmission
\submissiontrue % submission version
\newcommand{\name}{\textit{ARTist}\xspace}
\newcommand{\bname}{{\large\textbf{\textit{ARTist}\xspace}}}
\newcommand{\fullname}{\textit{ART Instrumentation and Security Toolkit}\xspace}
\newcommand{\compiler}{\textit{sec-compiler}\xspace}
\newcommand{\app}{\textit{deployment app}\xspace}
\newcommand{\taintlib}{\textit{TaintLib}\xspace}

\newcommand{\dexoat}{\textit{dex2oat}\xspace}
\newcommand{\bbdexoat}{{\large\textbf{\textit{DEX2OAT}\xspace}}~}
\newcommand{\bdexoat}{\textbf{\textit{DEX2OAT}\xspace}~}
\newcommand{\opt}{\textit{Optimizing}\xspace}
\newcommand{\quick}{\textit{Quick}\xspace}

\newcommand{\oat}{\textsf{oat}\xspace}

\newcommand{\dex}{\textsf{dex}\xspace}

\newcommand{\hins}{\textsf{HInstruction}\xspace}
\newcommand{\hinss}{\textsf{HInstructions}\xspace}
\newcommand{\hgraph}{\textsf{HGraph}\xspace}
\newcommand{\hformat}[1]{\textsf{#1}\xspace}

\newcommand{\cmark}[0]{\textcolor{green}{\ding{51}}}
\newcommand{\xmark}[0]{\textcolor{red}{\ding{55}}}

\newcommand{\instrapk}{\textsf{Instr}$_{\text{APK}}$\xspace}
\newcommand{\instroat}{\textsf{Instr}$_{\text{OAT}}$\xspace}
\newcommand{\instrdex}{\textsf{Instr}$_{\text{DEX}}$\xspace}
\newcommand{\instropt}{\textsf{Instr}$_{\text{OPT}}$\xspace}
\newcommand{\instrbin}{\textsf{Instr}$_{\text{BIN}}$\xspace}

\newcommand{\lsoone}{\textsf{LSO1}\xspace}
\newcommand{\lsotwo}{\textsf{LSO2}\xspace}
\newcommand{\lsothree}{\textsf{LSO3}\xspace}
\newcommand{\lsione}{\textsf{LSI1}\xspace}
\newcommand{\lsitwo}{\textsf{LSI2}\xspace}
\newcommand{\lsithree}{\textsf{LSI3}\xspace}

\documentclass{sig-alternate-05-2015}

\usepackage{etoolbox}
\patchcmd{\maketitle}{\@copyrightspace}{}{}{}

\renewcommand{\paragraph}[1]{\medskip\noindent\textbf{#1}}

\usepackage{etex}

\usepackage{color}
\usepackage{graphicx}
\usepackage[cmyk,table]{xcolor}

\usepackage{mdwlist}

\usepackage{lmodern}
\usepackage[utf8]{inputenc}
\usepackage[T1]{fontenc}
\usepackage[english]{babel}

\usepackage{microtype}  % Subliminal refinements towards typographical perfection
\usepackage{xspace}     % Adds a space behind a in-text macro unless the macro is followed by certain punctuation
\usepackage{multicol}   % allows switching between one and multicolumn format on the same page
\usepackage{balance}  % Balancing the Two Columns of Text on the Last Page

\usepackage{amssymb}   % symbols for natural numbers
\usepackage{amsfonts}  % symbols for natural numbers
\usepackage{amsmath}   % own operators
\usepackage{stmaryrd}  % The St Mary’s Road symbol font
\usepackage{MnSymbol}  % A Math Symbol Font

\usepackage{xtab}      % The xtab package enables long tables to be automatically broken at page boundaries
\usepackage{tabularx}  % Tables with a fixed total width and variable column width
\usepackage{multirow}  % table columns spanning multiple rows
\usepackage{booktabs}  % professional tables (advanced rules)

\usepackage{pdfpages}

\usepackage{float}

\definecolor{lightgray}{gray}{0.9}
\definecolor{gray}{gray}{0.8}
\definecolor{darkgray}{gray}{0.7}
\definecolor{lightblue}{rgb}{0.65,0.73,0.96}

\newcolumntype{a}{>{\columncolor{lightgray}}l}
\newcolumntype{b}{>{\columncolor{lightblue}}l}

\usepackage{framed}
\newfloat{otp}{htp}{loox}
\floatname{otp}{Output}

\newlength{\leftbarwidth}
\newlength{\leftbarsep}
\colorlet{leftbarcolor}{gray!70}

\usepackage{todonotes}
\usepackage{amssymb}% http://ctan.org/pkg/amssymb
\usepackage{pifont}% http://ctan.org/pkg/pifont
\usepackage{tikz}
\usepackage{pgf}
\usepackage{hhline}
\usepackage{listings}
\usepackage{paralist}
\usepackage{wasysym}
\usepackage{multirow}
\usepackage{etoolbox}

\usepackage[font=small,labelfont=bf,labelsep=colon]{caption}

\usepackage{enumitem}
\setlist[description]{leftmargin=0cm,labelindent=0cm}

\definecolor{circleblue}{RGB}{29,112,183}
\definecolor{circlegreen}{RGB}{0,141,54}

\newcommand*\circledblack[1]{\tikz[baseline=(char.base)]{
  \node[shape=circle,draw,inner sep=1pt,fill=black, text=white] (char) {#1};}}

\newcommand*\circledblue[1]{\tikz[baseline=(char.base)]{
  \node[shape=circle,draw,inner sep=1pt,fill=circleblue, text=white] (char) {#1};}}

\newcommand*\circledgreen[1]{\tikz[baseline=(char.base)]{
  \node[shape=circle,draw,inner sep=1pt,fill=circlegreen, text=white] (char) {#1};}}

\definecolor{Gray}{gray}{0.9}

\newcommand{\subparagraph}[1]{{\medskip\noindent\textbf{#1}}}

\definecolor{strings}{HTML}{007F3E} % comments
\definecolor{keywords}{HTML}{0000AE}
\date{}

\title{ARTist: The Android Runtime Instrumentation and Security Toolkit}

\usepackage{authblk}

\author[1]{\large Michael Backes}
\author[2]{Sven Bugiel}
\author[2]{Oliver Schranz}
\author[2]{Philipp von Styp-Rekowsky}
\author[2]{Sebastian Weisgerber}
\affil[1]{\small CISPA, Saarland University \& MPI-SWS, Saarland Informatics Campus}
\affil[2]{CISPA, Saarland University, Saarland Informatics Campus}

\begin{document}

\maketitle

\begin{abstract}

We present \name, a compiler-based application instrumentation solution for Android. \name is based on the new ART runtime and the on-device \dexoat compiler of Android, which replaced the interpreter-based managed runtime (DVM) from Android version 5 onwards. Since \dexoat is yet uncharted, our approach required first and foremost a thorough study of the compiler suite's internals and in particular of the new default compiler backend \opt. We document the results of this study in this paper to facilitate independent research on this topic and exemplify the viability of \name by realizing two use cases. Moreover, given that seminal works like TaintDroid hitherto depend on the now abandoned DVM, we conduct a case study on whether taint tracking can be re-instantiated using a compiler-based instrumentation framework. Overall, our results provide compelling arguments for preferring compiler-based instrumentation over alternative bytecode or binary rewriting approaches.

\end{abstract}

\section{Introduction}
\label{sec:introduction}
%!TEX root = ../paper.tex

Google's Android OS has become a popular subject of the security research community over the last few years. Among the different directions of research on improving Android's security, a dedicated line of work has successfully investigated how instrumentation of the interpreter (i.e., Dalvik virtual machine) can be leveraged for security purposes. This line of work comprises influencing works such as TaintDroid~\cite{EnGiBy_10:Taindroid} for analyzing privacy relevant data flows within applications, AppFence~\cite{HoHaJuScWe_2011:AppFence} for protecting the end-users' privacy, Moses~\cite{Russello:2012:MSO:2295136.2295140} for domain isolation, or Spandex~\cite{cox2014spandex} for password tracking, just to name a few.

However, with the release of Android 5 Lollipop, Google made a large technological leap by replacing the interpreter-based runtime with an on-device, ahead-of-time compilation of apps to platform specific bytecode that is executed in the new Android runtime (short ART). While this leap did not affect the app developers, it broke legacy compliance of all of the previously mentioned security solutions that rely on instrumentation of the DVM and it restricts them to Android versions prior to Lollipop. In fact, it has left the security research community with two choices for carrying on work that relies on instrumented runtimes: resorting to binary or bytecode rewriting techniques~\cite{DaSaKh_12:IARMDroid,Hao:2013:EAA:2484313.2484317} or adapting to the novel but uncharted on-device compiler infrastructure.

\pagebreak

\paragraph{Our contributions.} In this paper, we present a compiler-based solution that can not only be used to study the feasibility of re-instantiating previous solutions such as dynamic, intra-application taint tracking and dynamic permission enforcement, but, moreover, provides a more robust, reliable, and integrated application-layer instrumentation approach than previously possible. Concretely, we make the following contributions in this paper.

\textit{Study of the ART compiler suite.}  Since the novel ART compiler suite, \dexoat, is still uncharted, our solution required first and foremost a thorough study of the newly introduced \dexoat compiler. We provide, to the best of our knowledge, the first in-depth, comprehensive study of the internals of ART's compiler suite. In particular, we deep-dive into its most recent backend called \opt that became the default with Android 6 Marshmallow. Those new insights not only allow us to implement compiler-based security solutions, but also form expert knowledge that facilitates independent research on the topic.

\textit{Compiler-based app instrumentation.} We design and implement a novel approach, called \textit{ARTist} (ART Instrumentation and Security Toolkit), for application instrumentation based on an extended version of ART's on-device compiler \dexoat. Our system leverages the compiler's rich optimization framework to safely optimize the newly instrumented application code. The instrumentation process is guided by static analysis that utilizes the compiler's intermediate representation of the app's code as well as its static program information in order to efficiently determine instrumentation targets. A particular benefit of our solution, in contrast to alternative application layer solutions (i.e., bytecode or binary rewriting), is that the application signature is unchanged and therefore Android's signature-based same origin model and its central update utility remain intact. We thoroughly discuss further benefits and drawbacks of security-extended compilers on Android in comparison to bytecode and binary rewriting. Our results provide compelling arguments for preferring compiler-based instrumentation over alternative bytecode or binary rewriting approaches.

\textit{Feasibility study for compiler-based taint tracking.} To demonstrate the benefits of a solution such as our \textit{ARTist}, we conduct a case study 
on whether compiler-assisted instrumentation can be utilized to realize a dynamic intra-application taint tracking solution. Our resulting prototype is evaluated using microbenchmarks and its operational capability is shown using an open source test suite with known ground truth.

\paragraph{Outline.}  The remainder of this paper is structured as follows. In Section~\ref{sec:background}, we present the results of our study of the \dexoat compiler and its \opt backend. We analyze the requirements for an application-layer instrumentation solution in Section~\ref{sec:dilemma} and compare bytecode and binary rewriting with compiler-based approaches. We present our \name design in Section~\ref{sec:architecture}. Section~\ref{sec:usecases} illustrates use cases for \name, followed by a more detailed case study on compiler-assisted taint tracking. We discuss limitations and future work of our solution in Section~\ref{sec:discussion} and conclude this paper in Section~\ref{conclusion}.

%%% Local Variables:
%%% mode: latex
%%% TeX-master: "../paper"
%%% End:

\section{Background on ART and \bbdexoat}
\label{sec:background}
We provide general background information on Android's managed runtime to set the context of our compiler extensions (Section~\ref{sec:back:general}), and afterwards present technical background information on the compiler suite \dexoat (Section~\ref{sec:back:dex2oat}) and in particular on its \opt backend (Section~\ref{sec:back:optimizing}).

\subsection{Android Runtime}
\label{sec:back:general}

Android is essentially a Linux-based operating system with an extensive middleware software-stack on top of the kernel. The middleware provides native libraries, a feature-rich application framework that implements the Android SDK, and a managed runtime on top of which system as well as third-party applications and a small number of framework services are executed. The runtime executes bytecode generated from Java-based applications and Android's SDK components. The runtime provides the code executed within its environment the necessary hooks to interact with the rest of the system, such as the operating system, the application framework services, or the native Android user space (i.e., components running outside the managed runtime). Every process executing an application runtime environment is usually forked from a warmed-up process, called Zygote, which has all necessary libraries and a skeleton runtime for the app code preloaded.

\paragraph{Runtime prior to Android 5.}  On Android devices prior to version 5, the runtime consisted of the DEX bytecode interpreter (or Dalvik virtual machine), which was specifically designed for devices with constrained resources (e.g., register-based execution model instead of stack-based). It executes Dalvik executable bytecode (short \dex), which is created from the Java bytecode of applications at application build time. Thus, every application package ships the \dex bytecode compiled from the application Java sources. Additionally, since Android version 2.2, Dalvik uses just-in-time compilation of hotspot code segments in order to improve the runtime performance of applications.

\paragraph{Runtime since Android 5.}
With Android 5, Google moved over from an interpreter-based app execution to an on-device, ahead-of-time compilation of apps' \dex bytecode to native code that is executed in a newly introduced managed runtime called \textit{ART}. This shift in the runtime model was intended to address the app performance needs of Android's user and developer base. The new compiler suite was designed from scratch to allow for compile time optimizations that improve application performance, start up time, battery lifespan, and also to solve some well-known limitations of the previous interpreter-based runtime, such as the 65k method limit\footnote{\url{http://developer.android.com/tools/building/multidex.html}}. In particular, Google made the \textit{Optimizing} compiler backend, which was introduced as opt-in feature in Android 5, the default backend in Android 6. In the following Sections~\ref{sec:back:dex2oat} and \ref{sec:back:optimizing}, we will elaborate in more technical details on this new compiler suite and in particular on the \opt backend.

\paragraph{Prior documentation of ART.} Even though the Android source code is publicly available as a part of the Android Open Source Project~(AOSP), little attention has yet been given to ART from a security researcher's perspective. Paul Sabanal had an early look~\cite{hidingart} at the Android Runtime right after its silent introduction as a developer option on Android 4.4 KitKat. Beside providing information on the ART executable file formats, the paper discusses the idea of hiding rootkits in framework or app code, assuming root access has already been granted. However, especially in its early phase, the Android Runtime has undergone frequent changes, which by now make this documentation unfortunately outdated.\footnote{E.g., compare the documented \oat version 45 and the current version 63 at the time of this writing.}

 Another work~\cite{fuzzingart} focuses on fuzzing the new runtime with automatically generated input files in order to detect bugs and vulnerabilities. While it provides some high-level overview on the compiler structure and its backends, it unfortunately omits any deeper information on the \opt backend.

 Thus, this background section servers also the purpose of filling a gap in the technical documentation of those new Android features.

\begin{figure}
\includegraphics[width=\linewidth]{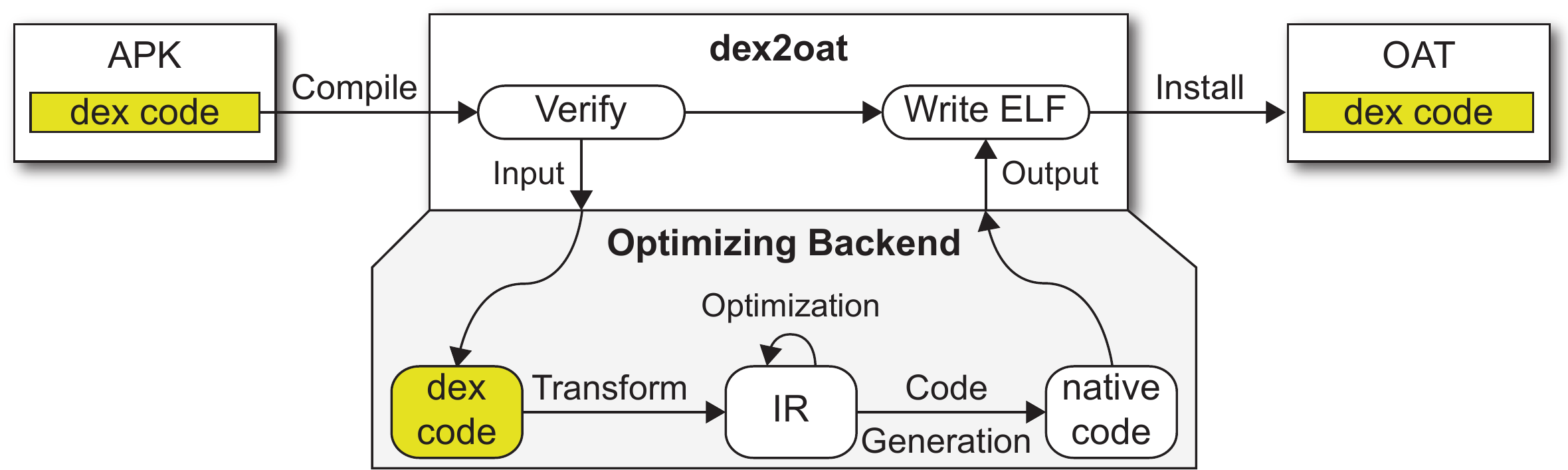}
\vspace*{+0.1cm}
\caption{A high-level overview of the \dexoat compiler using \opt backend including the transformation to the IR, optimizations, and native code generation. }
\label{figure:dex2oat}
\end{figure}

\subsection{\bdexoat Compiler Suite}
\label{sec:back:dex2oat}
Android's on-device compiler \dexoat is responsible for the validation of applications and their compilation to native code. It was designed from scratch to be highly flexible and of modular structure, providing numerous configuration possibilities, compiler backends, and native code generators for supported Android platforms. The general workflow of the compiler suite is depicted in Figure~\ref{figure:dex2oat} and its steps will be explained in the remainder of this section. Providing a full technically documentation of the entire compiler suite and all its intricacies would unfortunately exceed the space limitations of this paper. Therefore, in this section we only focus on the parts relevant to this paper.

\subsubsection{Input File Format}

As an input format, \dexoat expects the very same \dex files that DVM used to interpret.
This strategical decision ensured that neither developers nor app store operators needed to adapt their code to ART. Developers still upload their apps as Android Application Package (\textsf{APK}) files that bundle the app's code with its resources. When a new app is installed on the device, \dexoat compiles the app \dex bytecode and the ART runtime executes it, which is completely transparent for the end user. Using this strategy, ART is still compatible with the old Android app base without enforcing a fallback to interpretation, which would loose all benefits that the new compiler provides.

\subsubsection{Compilation}

Before the actual compilation is performed, each input \dex file is checked for validity. Those checks are more extensive and stricter than those implemented in the DVM in order to allow for state-of-the-art code optimizations. The compilation itself is done on a per-method base and can be
parallelized. \dexoat delegates the actual compilation completely to the backend and only writes the results of the compilation to an \textsf{oat} file along with the original \dex code. There are three compilation phases shared between all backends:
\begin{description}[font=\it]
\item[Transformation:] A graph-based intermediate representation (IR) is created from the \dex code. Depending on the actual backend, multiple IRs are possible.
\item[Optimization:] Given a populated IR graph, the code is optimized. Each backend provides its own set of optimization measures, ranging from very basic techniques to state-of-the-art algorithms.
\item[Native code generation:] The IR nodes are transformed to native code using a code generator for the specific CPU architecture of the current platform. The level of sophistication of the register allocation algorithm and implementation of the code generator depend on the backend.
\end{description}

\paragraph{Backends.} On an Android stock device running version 5 (Lollipop) or higher, \dexoat can choose between two different backends, \quick and \opt. Although \quick was \dexoat's default backend until Android 6, we focus in the remainder of this section and paper on the newer \opt backend. This choice is not only motivated by the fact that \opt is the default backend since Android 6 but also by the fact that \quick is essentially derived from Dalvik and lacks a sophisticated IR that can support state-of-the-art compiler optimizations---including sophisticated security-oriented algorithms. However, \opt is designed completely from scratch and little is yet known about its internal structure and design.
In Figure~\ref{figure:dex2oat} the compilation steps of \opt are depicted. More insights on the inner workings of the new default compiler backend will be provided in Section~\ref{sec:back:optimizing}.

\subsubsection{Oat File Format}
\textsf{Oat} files are Android’s new file format for apps that are ready to be loaded and executed by the ART runtime. Even though the format was newly created for the Android platform, technically speaking \oat files are specialized ELF shared objects that are loaded into processes, i.e., loading a compiled app into an application process is comparable to loading an (ELF) shared library into the process space of a dynamically linked executable. Besides the native code generated with \dexoat, \oat files contain the complete original \dex code, which is required to hold up consistency between the code that the developer wrote in Java, the \dex code that used to be interpreted, and the compiled code, or to allow fall back to interpretation mode when debugging apps.

\subsection{Optimizing Intermediate Representation}
\label{sec:back:optimizing}

\begin{figure}[t]
\centering
\includegraphics[width=.75\linewidth]{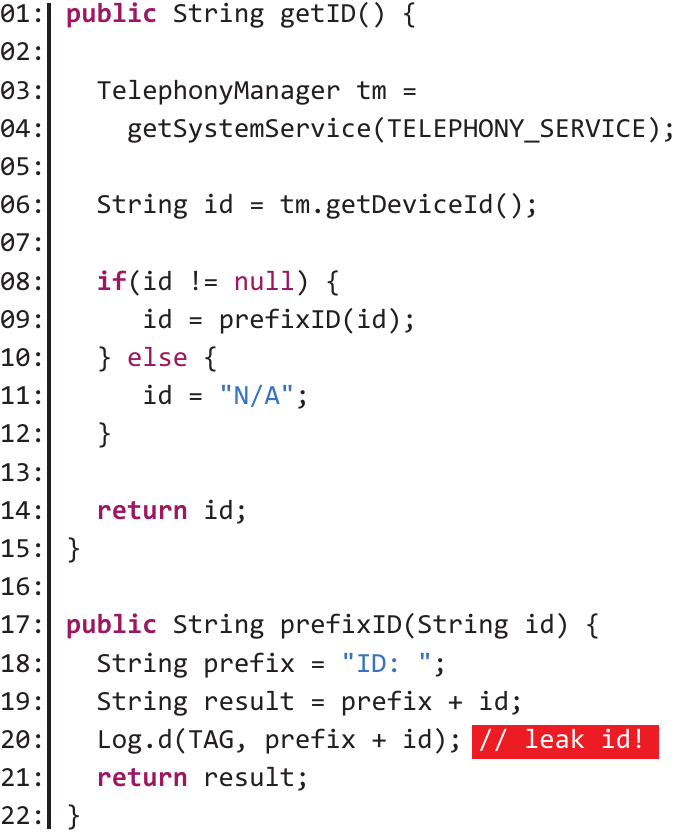}
\vspace*{+0.2cm}
\caption{An example code snippet containing a leak of the device's phone number to the logging facility.}
\label{figure:codesnippet}
\end{figure}

\begin{figure}[t]
\centering
\includegraphics[width=.8\linewidth]{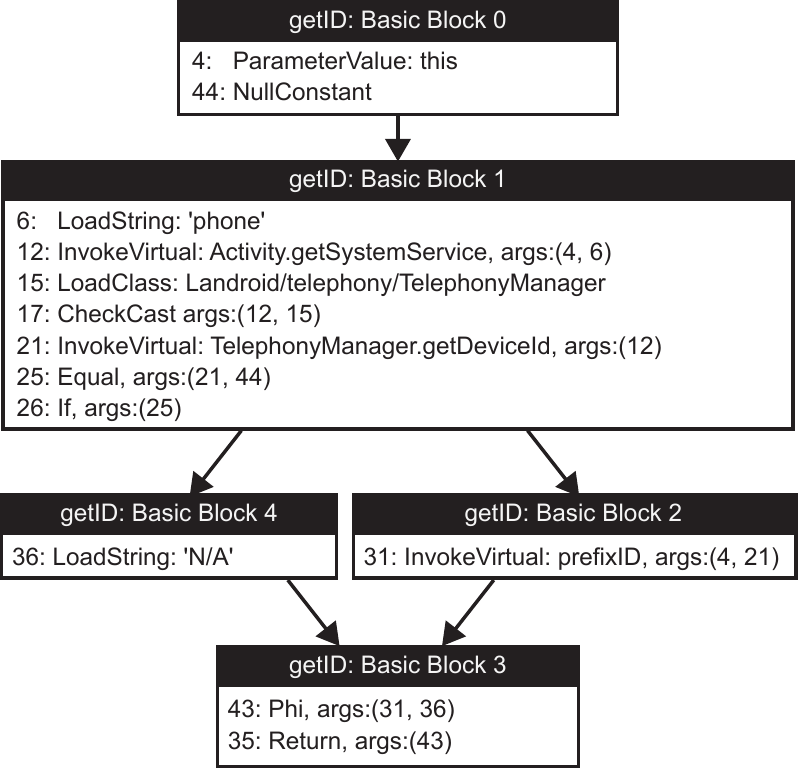}
\vspace{+0.2cm}
\caption{Generated IR in SSA form for the \texttt{getID()} method in Figure~\ref{figure:codesnippet}.}
\label{figure:ir}
\end{figure}

We introduce insights into \dexoat's \opt backend, which we derived mainly from the AOSP source code of the ART project's master branch between June 2015 and February 2016. \opt's intermediate representation is essentially a control flow graph on the method level, which the Android developers denote as \hgraph. It is further enriched with structural data about the program and populated with instruction nodes, denoted as \hinss. Figure~\ref{figure:codesnippet} presents an example Java code and Figure~\ref{figure:ir} presents the resulting\footnote{Presented code is simplified and limited to relevant instructions for the sake of readability.} \hgraph of the \texttt{getID} function in the \opt IR. We will come back to this example in our case study in Section~\ref{sec:usecases:casestudy}.

\subsubsection{\hgraph}
The \hgraph serves as the single intermediate representation of the app code. When the graph is created, \dex instructions of the app's bytecode are scanned one by one and the corresponding \hinss are created and interlinked with the current basic block and the graph. In order to allow for complex optimizations, the graph is transformed into a \textit{single static assignment form}~(SSA). Pairs of value definitions and usage, so-called \textit{def-use-pairs}, are created during a liveness analysis and explicitly interlinked afterwards. At this point, \textit{phi nodes} are introduced where static analysis cannot reliably decide which value will be assigned at a given position.

 In this form, the graph is amenable to a multitude of possible optimizations. The available optimizations includes algorithms such as \textit{BoundsCheckElimination} to remove redundant bounds checks, \textit{GVNOptimization}  to remove duplicate code, dead code elimination, or \textit{loop invariant code motion} to optimize hotspot code in loops. In the following Section~\ref{sec:architecture}, we will show how this form is also amenable to security-oriented instrumentation, thus supporting compiler-based security solutions on Android, such as dynamic taint tracking (see Section~\ref{sec:usecases:casestudy}).

\subsubsection{\hinss}
The \hgraph nodes roughly correspond to \dex instructions. The supplementary material provides an overview of the \dex instructions and their corresponding \hinss counterparts. Beside this transformation, nodes in the \hgraph have additional attributes that have no equivalent in \dex bytecode (e.g., an SSA index). The \hinss distinguish between arguments and inputs. While the former corresponds to the arguments given to an operator or method, the latter encodes additional dependencies that may not be immediately observable given only the underlying \dex code, as in the case of static method invocations that in addition to their arguments have an \hformat{HLoadClass} or \hformat{HClinit} as their input. All \hinss share a basic set of information: Type, inputs, uses, id, and further data is attached to each node in order to ease the creation of and working with the \hgraph. Each node is uniquely identified within the graph by its id that is assigned and incremented continuously during node creation. The type can be \textsf{void} for methods that have no return value, \textsf{not} for strings and object types of any kind, and additionally any of the Java primitive types. In order to get the actual object type, a fallback to the original \dex file is required. This loose coupling between \hinss and \dex instructions as well as the presence of a method local \dex program counter in each node show that the IR nodes are not completely independent of the original \dex file.

\paragraph{Semantic consistency.} In addition to the instructions that represent the original application logic, the \hgraph also contains meta-instructions to preserve the semantic consistency between the original Java code of the developer, the \dex bytecode shipped with APKs, and the native bytecode actually executed in ART. First, additional instructions are inlined into the graph to support meaningful debugging (e.g., to map from segmentation faults in ART to actual stack traces) and to conduct various forms of runtime checks (e.g., checking type casting, bounds checking, division-by-zero checks, or null pointer exceptions). Second, instructions to represent so-called suspension points are added, which effectively subdivide the application code into multiple chunks. Each suspension point between two chunks acts as a synchronization point between native code and original \dex bytecode in the program execution and also serves as an entry point for garbage collectors or debuggers.

%%% Local Variables:
%%% mode: latex
%%% TeX-master: "../paper"
%%% End:

\section{The Case for Compiler-Assisted Security on Android}
\label{sec:dilemma}
\begin{figure}[t]
\includegraphics[width=\linewidth]{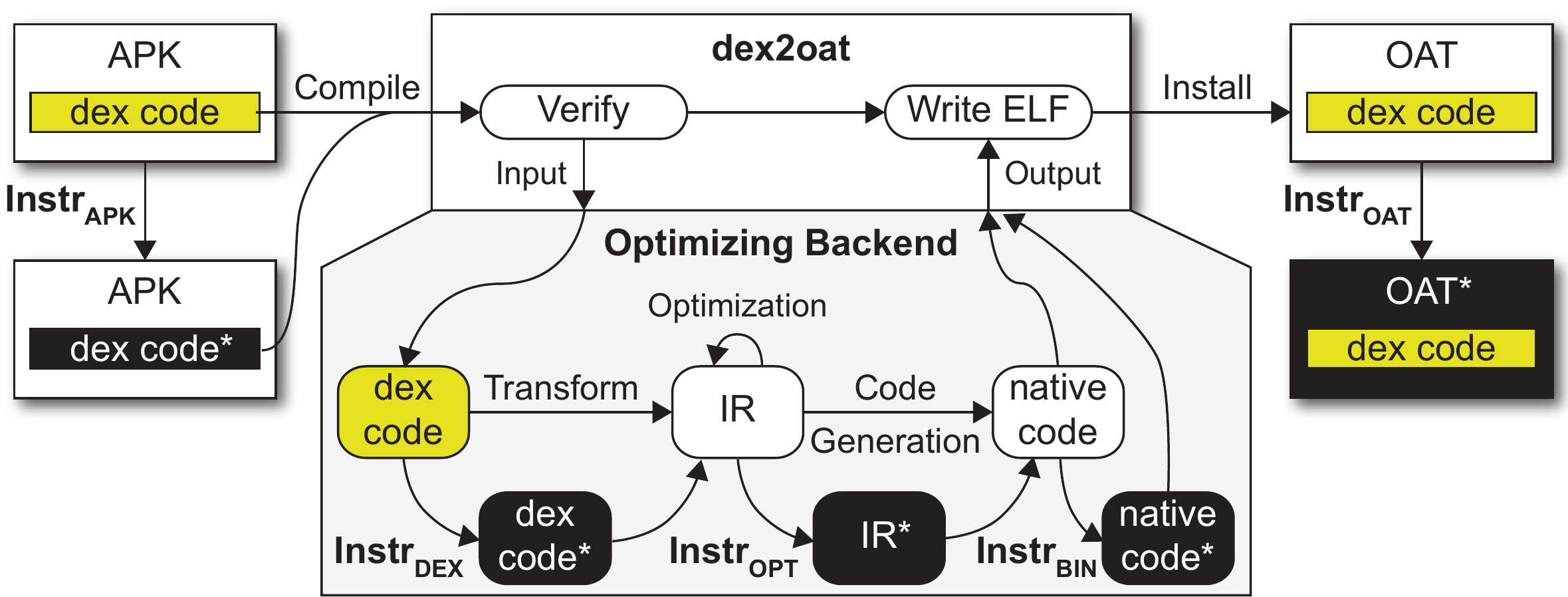}
\vspace{+0.1cm}
\caption{The code instrumentation points before, during, and after the compilation for different representations of the app code. Instrumented code is depicted in black boxes.}
\label{figure:dex2oat3}
\end{figure}

A dedicated line of work, including the TaintDroid project and its derivatives (such as~\cite{HoHaJuScWe_2011:AppFence, droidbox}), relied on instrumentation of the now abandoned Dalvik virtual machine. As a consequence, the research community faces the dilemma on how to continue this line of work and is left with two choices (see Figure~\ref{figure:dex2oat3}): Either compensating the missing runtime instrumentation through app rewriting techniques---\dex bytecode (\instrapk) or binary (\instroat)---or taking advantage of Android's new compiler suite (\instrdex, \instropt, and \instrbin). Although \dex bytecode rewriting is well-established in contexts such as inline reference monitoring~\cite{JeMiVa_11:DrAndroid, DaSaKh_12:IARMDroid, Davis:2013,backes13TACAS} and taint analysis~\cite{schutte2014appcaulk}, and ART now supports porting binary rewriting techniques from commodity systems, we build in this paper on compiler-based instrumentation to not only re-instantiate previous approaches that relied on Dalvik VM instrumentation, but also to explore novel security solutions that leverage the compiler features.

In the following, we analyze the concrete requirements that an instrumentation solution should provide and discuss for each of the above approaches (i.e., bytecode rewriting \instrapk, binary rewriting \instroat, and compiler-based instrumentation) their respective benefits and shortcomings in fulfilling those requirements. Table~\ref{table:InstrumentationComparison} provides a summary of our requirements analysis.

\begin{table}
\footnotesize
%\resizebox{\columnwidth}{!}{
\begin{tabular}{|p{2.6cm}|c|c|c| } 
\cline{2-4}
\multicolumn{1}{c|}{}     & \shortstack{Bytecode\\rewriting} & \shortstack{Compiler-\\based} & \shortstack{Binary\\rewriting} \\
\hline
\textbf{R1.} Enforceable security policies  & \multicolumn{3}{|c|}{identical} \\
\hline
\textbf{R2.} Strong security boundary & \xmark & \xmark & \xmark \\
\hline
\textbf{R3.} Application layer only & \cmark & \cmark & \cmark \\
\hline
\textbf{R4.} User privilege only & \cmark & (\xmark) & (\xmark) \\
\hline
\textbf{R5.} Platform independence & \cmark & \cmark & \xmark \\
\hline
\textbf{R6.} Signature preservation& \xmark & \cmark & \cmark \\
\hline
\textbf{R7.} Robustness against optimization & \xmark & \cmark & \cmark \\
\hline
\textbf{R8.} Integrated approach & \xmark & \cmark & \xmark \\
\hline
% Merged with previous
%Reusable program information & \multirow{2}{*}{\xmark} & \multirow{2}{*}{\cmark} & \multirow{2}{*}{\xmark} \\
%\hline
\textbf{R9.} Supported versions& all & 6+ & 5+ \\
% \hline
% \multicolumn{4}{|l|}{\textcolor{red}{*} While compiler-based instrumentation is possible } \\
% \multicolumn{4}{|l|}{\;\;\;since Android 5, \opt is only available from} \\
% \multicolumn{4}{|l|}{\;\;\;Android 6.} \\ 
\hline
\multicolumn{4}{c}{\cmark = fulfilled; \xmark = not fulfilled}
\end{tabular}
%} 
\vspace{+0.2cm}
\caption{Comparison of security and deployment features between bytecode rewriting, compiler-based instrumentation, and binary rewriting. }
\label{table:InstrumentationComparison}
\end{table}

\paragraph{R1. Enforceable security policies.} Each of the three approaches operates on one of the different representations of the same app code, i.e., bytecode, IR, or binary. Hence, all three approaches are identical in their capabilities of instrumenting the code and none of the solutions addresses any security policy alone.

\paragraph{R2. Strong security boundary.} Both rewriting and the compiler-based approach rely on injecting monitoring code into the app's process space and can therefore not provide a strong security boundary between monitoring and (potentially) malicious app code (\xmark), e.g., native code. Thus, all of them can only provide security guarantees for at most honest-but-curious apps.

\paragraph{R3. Application layer only.} All approaches can be implemented purely at application layer (\cmark). Deploying bytecode rewriting techniques~\instrapk in form of separate apps has been presented in the literature~\cite{JeMiVa_11:DrAndroid, DaSaKh_12:IARMDroid, Davis:2013,backes13TACAS}. A compiler-based solution can be deployed as a separate app that ships and controls the security-instrumented compiler suite (see also Section~\ref{sec:architecture:app}). For both, the compiler-based approach and the binary rewriting, the main requirement is access to the storage location of applications' \oat files, which does not require system modification.

\paragraph{R4. User privilege only.} While \dex code is freely available for non-forward locked apps, accessing applications' \oat files makes elevated privileges necessary. However, in Section \ref{sec:discussion:artist:impllimit} we discuss an approach that would allow both, binary and compiler-based rewriting, to circumvent this problem without requiring elevated privileges. 

\paragraph{R5. Platform independence.} Bytecode rewriting \instrapk (\cmark) and compiler-based instrumentation (\cmark) can be applied on all platforms supported by Android, since they modify the code before platform-dependent native code is generated. Binary rewriting \instroat, in contrast, depends on the actual hardware architecture of the platform (\xmark).

\paragraph{R6. App signature preservation.} App signatures are the foundation of Android's same origin model that governs the app update policy or sharing of resources between apps, like a common process or UID. Consequently, modifying bytecode \instrapk and the resulting obligation to resign and repackage apps breaks this same origin model (\xmark). In contrast, compiler-based instrumentation (\cmark) and binary rewriting \instroat (\cmark) do not modify the original app package and therefore do not invalidate the signature.

\paragraph{R7. Robustness against code optimization.} The instrumentation point determines whether any instrumented code will be subject to optimization at compile time. Applying optimization algorithms to instrumented code has the potential to interfere with the semantics of the modification through, e.g., instruction reordering, inlining, or similar techniques of state-of-the-art compilers like \opt. Current bytecode rewriting approaches \instrapk are applied before compilation, thus any instrumentation has to be robust against optimization by \opt---an aspect not yet further investigated by contemporary research (\xmark). On the other hand, binary rewriting \instroat is restricted to instrumenting optimized code (\cmark), but misses the chance to reuse the rich \opt framework to also optimize added security code. The sweet spot is compiler-based instrumentation that provides full control over which optimizations are applied when and in which ordering (\cmark). Even more, this enables creating optimizations that are specifically tailored towards improving the instrumented code by utilizing the static program information that is present in the compiler.

\paragraph{R8. Integration into toolchain.} Integrating an instrumentation system into an existing toolchain ensures perpetual development and fixes by the community as well as access to established and well-tested tools and frameworks. In this case, even though the ART project is open source and therefore open to the community, the compiler is mostly maintained by Google itself. Consequently, compiler-driven solutions that do not break with the toolchain's regular functionality, benefit from the continuing improvements (\cmark). In the case of \name, the amount of code that needed to be changed is minimal and therefore easy to adapt for newer versions of the toolchain. Bytecode rewriting \instrapk and binary rewriting \instroat are developed separately from the toolchain and do not reap those benefits (\xmark).

\paragraph{R9. Version support.} While bytecode instrumentation \instrapk can be applied to \textit{all} Android versions, compiler-based approaches and binary rewriting \instroat depend on ART and therefore can only be applied since Lollipop (\textit{5+}), where a compiler-based solution (as presented here) should utilize the \opt backend on Android \textit{6+} in preference to \textit{Quick}.

\paragraph{Sweet spot.} In conclusion, comparing the security and deployment features that the three available instrumentation approaches provide, a compiler-based approach has very appealing properties and occupies a sweet spot among all approaches.

%%% Local Variables:
%%% mode: latex
%%% TeX-master: "../paper"
%%% End:

\section{\bname~Design}
\label{sec:architecture}

In this section, we present the architecture of the \fullname. \name consists of two separate components: a security-instrumented compiler (\compiler) and an app to deploy the compiler (\app). The \compiler is our implementation of a compile-time instrumentation tool that is based on the \dexoat compiler. The latter is a regular Android application that ships, deploys, and manages the \compiler.

\subsection{Security-Instrumented Compiler}

\paragraph{Choice of instrumentation point.} The general concept of security-instrumented compilers is not restricted in its modifications of the compilation code. Given \dexoat's modular design, there are immediately multiple possibilities apparent where app modifying code could be placed. For instance, \dexoat's design would easily allow porting bytecode and binary rewriting approaches (\instrdex \& \instrbin) into the compiler infrastructure (cf.~Figure~\ref{figure:dex2oat3}). Of the different choices, \name's \compiler is concretely designed to operate on the intermediate representation of \dexoat's \opt backend (\instropt), where the existing optimization infrastructure and static code information in the \opt IR allow for efficient and precise code modification. More precisely, our app instrumentation code is realized as an \hformat{HOptimization}, which, as a result, is no different than other optimizations in the sense that they are provided with required information, such as the current method's \hformat{HGraph}, and are modularly integrated into the optimization workflow. As \hformat{HOptimization}, our security instrumentation logic has full control over the ordering and execution of optimizations, which opens up the opportunity to optimize the already instrumented code by creating or applying compatible optimizations that improve the performance of the security code.

Generally, using the \hformat{HOptimization} interface one can extend the compiler with custom functionality that is decoupled from \dexoat's code base. We will refer to those independent extensions as \textit{Modules} for the remainder of this paper.

\paragraph{Spotting instrumentation targets.} \hformat{HGraph} supports the visitor pattern~\cite{Gamma:1995:DPE:186897} that enables us to iterate over, inspect, and modify each single \hformat{HInstruction} of the app's code. In contrast to method hooking techniques, we can therefore operate at the instruction level. In \name, \hformat{HGraphVisitors} are primarily used to identify instrumentation targets and apply the desired modification. However, they can also be utilized to bootstrap static analysis. We will see concrete implementations using a visitor to collect instrumentation sites for our dynamic permission enforcement system in Section~\ref{sec:usecases:IRM} and starting points for backward slicing in our taint tracking case study in Section~\ref{sec:usecases:casestudy}.

\paragraph{Modification capabilities.} Given an instrumentation target in the form of an \hins, there are several possibilities for modification like changing types, inputs, or even removing/replacing the instruction altogether. It is also possible to instantiate new instructions of any type and inline them before or after the current target. \name provides a new API dedicated to automate such modifications if provided with the information which methods, where and what to instrument. Since we are only generating nodes in the form of \hins objects and insert them into the \hgraph, we do not have to modify the generation of native code from the IR. This means that the code generator is agnostic towards our changes and receives no unexpected instructions or structures. The result of this integrated solution is that we still take advantage of the robustness of \opt's code generators, which are well-tested, constantly improved, and in productive use on every stock Android phone running version 6+.

\paragraph{Configuration.} Using \textit{Modules}, the instrumentation and modification process is already flexible. To further increase flexibility, \textit{Modules} can, in turn, depend on policy configuration files that govern the instrumentation process. While the design of such policy files highly depends on the concrete \textit{Module}, there are recurring and common patterns, for instance, the amount and type of instrumentation targets that should be detected, as demonstrated in Section~\ref{sec:usecases:IRM}. In general, this allows adaptation of existing instrumentation solutions to new targets or provisioning them with new security policies.

\subsection{Compiler Deployment App}
\label{sec:architecture:app}

Responsibility of the \app is to deploy the \compiler at application layer in addition to the system's \dexoat binary. Using \app, one can create security-instrumented versions of installed applications by re-compiling the apps' bytecode with \compiler and replacing the \oat files stored\footnote{Located at \path{/data/app/<package-name>-1/oat/arm/base.odex}} on filesystem. To make the Android runtime agnostic to this instrumentation, two particular challenges had to be overcome. First, Android has mechanisms in place to verify that \oat files correspond to their respective apps and that the paths of the \oat files are correct. Our implementation solves this challenge by rewriting paths and checksums to match those that the system \textit{dex2oat} would have generated. Second, the \oat files are by default stored at and loaded from a protected location to which 3$^{rd}$ party apps have no access. A na{\"i}ve solution to this problem would be to require extended privileges for our \app (e.g., a dedicated SELinux type or root on security-relaxed after-market ROMs). We discuss alternatives to the na{\"i}ve approach in Section~\ref{sec:discussion}, which abstain from extended privileges by using app virtualization or reference hijacking. After solving those challenges, the Android default runtime will load the instrumented \oat file while remaining agnostic to the fact that we replaced it.

\paragraph{Executing the compiler.} Instead of shipping \app with a statically linked \dexoat binary that includes our \name extensions, we opted for utilizing a copy of Android's default \dexoat binary and leveraging its modularity to ship our extensions to the compiler suite as separate libraries. We use the \textsf{LD\_LIBRARY\_PATH} environment variable to ensure that our \dexoat loads and dynamically links our \name libraries, such as \textsf{libart-compiler.so}, from the assets directory of the \app.

\paragraph{Inlining custom code.} While the instrumentation with \name already provides powerful tools to modify the application, most security solutions require an additional custom code library within the app (e.g., additional taint tracking logic in Section~\ref{sec:usecases:casestudy}). To facilitate adding custom code to an instrumented app, \app has a preprocessing step that is executed before the app's bytecode is compiled. This step utilizes the \textsf{DexMerger} utility to combine the app's original bytecode with the additional code library. During compilation, connections between original and new code are built in form of invocations of the added code's methods.

%%% Local Variables: 
%%% mode: latex
%%% TeX-master: "../paper"
%%% End: 

\section{Use Cases}
\label{sec:usecases}
We demonstrate the applicability and usefulness of our system by discussing several use cases out of which we  exemplarily realized two as \name \textit{Modules}. First, we implemented an Inline Reference Monitor (IRM) injection \textit{Module} to allow for dynamic permission enforcement. Second, we conduct a case study on realizing intra-app taint tracking through inlining of taint tracking code. In addition, we discuss further ideas for \name \textit{Modules}.

\newpage
\subsection{IRM for Dynamic Permission Enforcement}
\label{sec:usecases:IRM}

In the literature, Inline Reference Monitoring (IRM) is mostly implemented by modifying the bytecode before the installation~\cite{JeMiVa_11:DrAndroid,DaSaKh_12:IARMDroid}~or by hooking into an application's method at the caller or callee side at runtime~\cite{backes13TACAS}. By utilizing a security-instrumented compiler, IRM can be implemented without the need to resign and repackage apps as it is required by established approaches. Moreover, \dexoat-based IRM can operate at instruction granularity instead of at the method level. Those capabilities are showcased by our IRM injection \textit{module} that allows for dynamic permission enforcement, as shown by~\cite{backes13TACAS,JeMiVa_11:DrAndroid,XuSaAn_12:Aurasium} on Android versions before Marshmallow.

The module is split into two distinct parts, the code injection routine that will inline permission enforcement code and the accompanying library that acts as a policy decision point. While the first directs the instrumentation process at installation time, the latter enforces the user's policy at runtime. 

\paragraph{Code injection.} We first utilize \name to locate the call sites of permission-protected SDK methods that are defined in a policy configuration file. Afterwards, \name injects additional calls to our companioning library right before the call sites to check whether the critical method invocations should be allowed. This ensures that the control flow is diverted to our policy decision point before the execution of permission-protected methods. 

The limitations imposed by the choice of this rather basic strategy are discussed in Section \ref{sec:discussion:permmodule:limit}.

\paragraph{Policy decision point.} The library that our \textit{Module} injects into target apps provides methods to check their current state of permissions. Based on the given user permission policy, the library either allows or rejects the execution of a protected SDK method. 

\subsection{Case study: Taint Tracking}
\label{sec:usecases:casestudy}

Established approaches for dynamic taint tracking on Android~\cite{EnGiBy_10:Taindroid} rely on instrumenting the by now scrapped DVM for intra-application taint tracking or directly rewrite bytecode \cite{schutte2014appcaulk}. In this case study, we explore the applicability of a compiler-based instrumentation framework like \name to re-instantiate intra-app taint tracking for applications on Android version 6 and higher. That is, through a prototypical implementation, we want to investigate whether inlining taint tracking logic into the application code base with \name at compilation time can be a surrogate for solutions prior to Android version 5. Please note, that this case study does not aim at a full replacement of existing solutions like TaintDroid~\cite{EnGiBy_10:Taindroid}, but at demonstrating a new potential foundation for future taint-tracking on Android.

\subsubsection{Module Design}
\label{sec:usecase:arch}

In general, we want to track information as it flows through the code using tracking logic inlined by a new \hformat{HOptimization} in the \opt backend. However, simply assigning each single value that should be tracked a taint tag and updating the tag for each single instruction operating on it will incur a major performance penalty. To minimalize the runtime impact, we split our approach into two phases: \textit{analysis} and \textit{instrumentation}. In the \textit{analysis} phase, we identify flows of tainted information between \textit{sources} and \textit{sinks}. By restricting ourselves only to those relevant flows of the values we are interested in, we avoid generating irrelevant but costly taint tracking code for parts of the method that never actually influence the data that is obvserved and gain noticable performance improvements over more na{\"i}ve taint tracking. In the \textit{instrumentation} phase, code will be inlined that creates, propagates, and checks the taint values along the identified data flows. Our combined analysis and instrumentation achieves flow-, path-, object-, and context-sensitive taint tracking.

While \cite{schutte2014appcaulk} and \cite{livshits2013towards} also utilize static analysis to optimize and guide the instrumentation process, both assume a holistic view on the application in form of a control or data flow graph. In contrast, \dexoat backends operate on a per-method level, leaving the primary challenge for our taint tracking \textit{Module} to achieve similar tracking properties while inspecting one method at a time. 
A na{\"i}ve solution to this problem would be to retrofit the compiler suite to provide an application-wide view and instrumentation. However, our prototype demonstrates how we can still achieve taint tracking for the whole application while restricting ourselves to a per-method view and instrumentation. To this end, we introduce in the following a new design for storing and propagating taint tags, in particular we have to refine the definitions of \textit{sink} and \textit{source}.

\subsubsection{Analysis Phase}
\label{sec:usecase:analphase}

In order to optimize the instrumentation with taint tracking code, we exploit the processing features (e.g., \hgraph's Visitor~\cite{Gamma:1995:DPE:186897} pattern support) of the \dexoat compiler to detect the data flow sources and sinks and afterwards use its static analysis features to identify the relevant data flows and the operations along those flows that have to be instrumented.

\subparagraph{Refining source and sink definition.} The literature on taint tracking for Android defines sources and sinks as the API methods that input privacy-sensitive information into the application process (e.g., framework functions that return sensitive data, such as the location or telephony API) or, respectively, leak privacy-sensitive information from the application process (e.g., file handles, Internet sockets, or logging facilities). Since \dexoat is operating on a per-method level, we cannot assume that our analysis is able to always connect a sink and a source (e.g., when they are located in different methods). To address this problem, we have to connect the data flows of tainted variables across the different methods while maintaining the per-method-based analysis. To this end, we introduce in addition to the above mentioned sinks and sources from the literature---in the following denoted as \textit{global sinks/sources}---new \textit{method-local sinks/sources}, more precisely \hinss, which form the entry and exit points for inter-procedural data flows. Thus, global sinks and sources are points of interest for taint tag creation and check, respectively, while local sinks and sources are for inter-procedural tag propagation. For local sinks and sources we differentiate between three categories each: \textit{local sources} include arguments provided to the current method~(\lsoone), return values from method invocations~(\lsotwo), and values read from fields~(\lsothree). Conversely, \textit{local sinks} are method invocations that leak values through its parameters from the current method~(\lsione), return statements of the current method~(\lsitwo), and field setting instructions~(\lsithree). At the beginning of the analysis phase we collect all sinks within all methods and in a subsequent step detect all relevant sources for those sinks (see next paragraph).

\begin{figure*}[t]
  \centering
  \includegraphics[width=\linewidth]{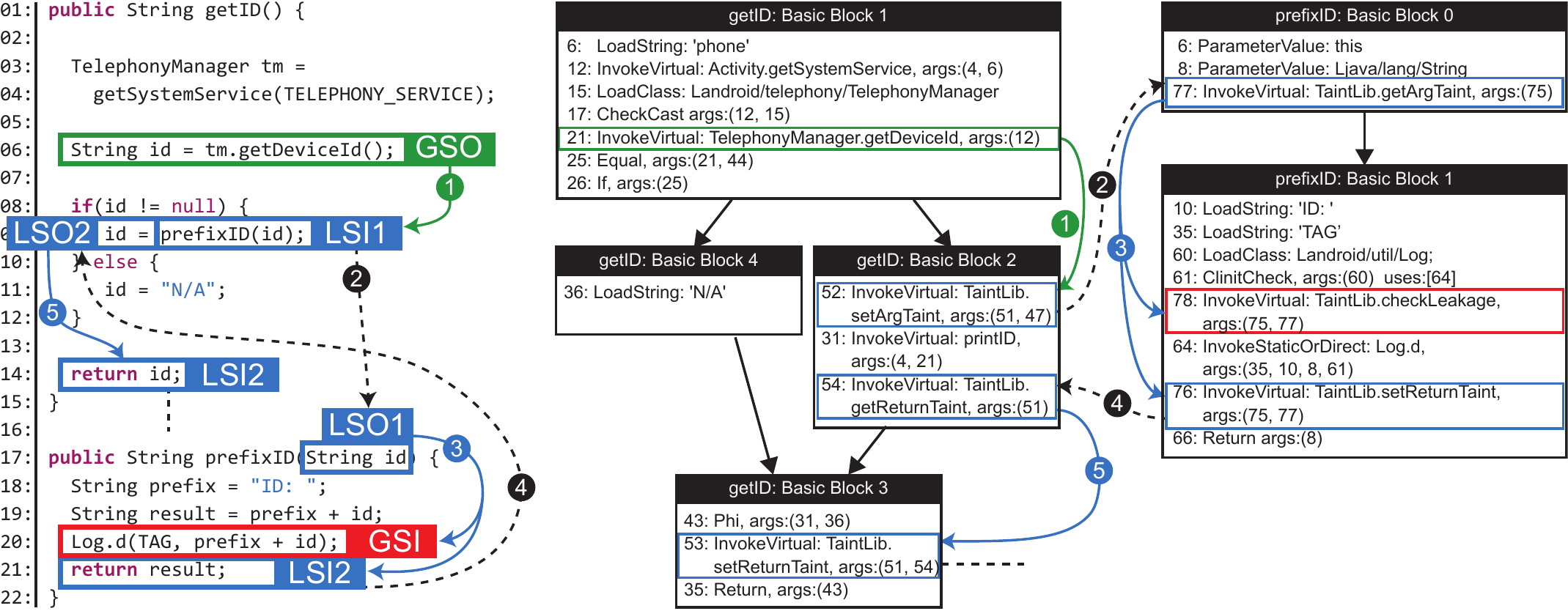}
  \vspace{+0.1cm}
  \caption{Tracking tainted variable \textit{id} from example of Figure~\ref{figure:codesnippet}. All discovered sinks and sources are marked. Solid lines indicate intra-procedural data flows of tainted variables, dashed lines inter-procedural data flows between local sink-source pairs. Right hand side depicts the inlined taint tracking code to propagate taint tags.}
  \label{fig:tracking}
\end{figure*}

\subparagraph{Creating intra-procedural data flows.} After collecting all local and global sinks, we create for each sink a backward slice within the currently analyzed method by inspecting the current instruction and recursively tracing back its input parameters depending on the concrete instruction type to discover all sources that influence this sink. For instance, Figure~\ref{fig:tracking} continues the example code from Figure~\ref{figure:codesnippet}. In the Java code on the left hand side, all sinks have been identified (i.e., the parameter \texttt{id} passed to function \texttt{prefixID} in line~9 is a local sink of type \lsione, the return statements in lines~14 and 21 form local sinks of type \lsitwo, and in line~20 the call to \texttt{Log} forms a global sink). Using backwards slicing (solid lines \circledgreen{1}, \circledblue{3}, and \circledblue{5}) the local sources in lines~17 (\lsoone) and 9 (\lsotwo) as well as the global source in line 6 (\texttt{getDeviceID} call to retrieve device's phone number) have been identified. Each resulting backward slice is defined by its starting point (i.e., the sink) and all found endpoints (i.e., the sources). Constants cannot be tainted and are therefore explicitly omitted as sources. Together those backward slices form the input for the instrumentation phase. 

Note that high precision of this analysis is desirable but \textit{not} required for secure taint tracking. Higher precision of this backwards slicing 
%(e.g., path sensitivity) 
helps in removing irrelevant taint tracking code for the observed values and hence improves performance of the instrumented application's process, but actual taint propagation occurs at \textit{runtime} and the set of our slices contains a superset\footnote{The use of reflection is an exception to this rule since this poses serious problems for static analyses.} of the relevant flows for complete tracking of tainted values.
%\TODO{anything else to say? is this parameter sensitive, path sensitive etc?}

\subsubsection{Instrumentation Phase}
\label{sec:usecase:analphase}

In the instrumentation phase, we inline code that at runtime creates taint tags for global sources and that checks taints at global sinks. Additionally, we inline code that inter-procedurally propagates taints at runtime from a local sink to a local source, ensuring the data flow of a tainted value across multiple methods correctly propagates the taints.

\subparagraph{TaintLib.} In order to improve the flexibility of our solution by not restricting our system to a specific implementation for storing, updating, and checking taints, we make use of \name's modular design and deploy the instrumentation code in form of a new \textit{companion library} called \taintlib that is merged by \name into the app code at compile time. \taintlib in turn relies on a policy file that defines the global and local sources/sinks as well as the sources' taints tags. \taintlib provides source-type-specific \textit{taint-set} methods, calls to which are inlined at all sources, and sink-type-specific \textit{taint-get} methods, calls to which are inlined at all sinks. By injecting \taintlib method calls instead of concrete taint tracking logic, we decouple the instrumentation from the taint management code. For global sources, \textit{taint-get} retrieves and sets the taint tag according to the policy and \textit{taint-set} at global sinks checks\footnote{While a na{\"i}ve check halts the program when tainted data is about to leak, invoking a sanitizer as suggested by \cite{livshits2013towards} can easily be implemented in \name.} the taint tag. In contrast, for local sinks \textit{taint-set} propagates the tag together with the tainted value to the next local source, where it is retrieved with \textit{taint-get}. By instrumenting all methods alike, an implicit contract between all methods is established and fulfilled, i.e., every time a \textit{taint-get} tries to obtain the taint value of a method parameter on the callee side, we assume that the corresponding \textit{taint-set} has been executed in the calling method to provide the taint data. In case the slice contains multiple \textit{sources}, the output of their corresponding \textit{taint-gets} is combined by injecting a call to a combination method that will return the merged taint tag.

To continue our running example, the right hand side of Figure~\ref{fig:tracking} presents the IR of the code snippet with \textit{taint-set} and \textit{taint-get} calls inlined. For instance, the \texttt{setArgTaint} call for \lsione in basic block~2 of \texttt{getID} (\hinss~52) precedes the local sink in \hinss~31 that invokes the \texttt{prefixID} function. The \texttt{setArgTaint} instruction transfers the taint of \texttt{id} inter-procedurally to the \texttt{getArgTaint} instruction in \hins~77 of basic block~0 of \texttt{prefixID} (dashed line \circledblack{2}), from where it is intra-procedurally propagated using the backwards slicing information (solid line \circledblue{3}). Similarly, the taint is propagated back from \texttt{prefixID} to \texttt{getID} through the return statement and variable assignment (dashed line \circledblack{4}).

\subparagraph{Inter-procedural taint tag propagation channel.} In the case of parameters (\lsoone and \lsione) and method returns (\lsotwo and \lsitwo), there are at runtime always pairs of \textit{taint-sets} and \textit{taint-gets}, given by the fact that for each callee method, there is a caller method that also has been instrumented. Combining this with the observation that a caller-callee method pair is always executed by the same thread, the taint propagation can be realized using \textsf{thread local storage} for a \textsf{taint stack}. At the caller side, the taint information is \textsf{pushed} onto a per-thread stack and at the callee side it is \textsf{popped} again, vaguely resembling the x86 calling convention for passing arguments to methods. Keeping in mind that almost every injected \textit{TaintLib} method call accesses the taint information, replacing more straightforward approaches for taint storage (like a single \textsf{HashMap}) with cheaper stack operations also benefits the overall performance of our taint tracking solution.

In the case of field operations (\lsothree and \lsithree), we can neither assume them to appear in pairs nor to be executed on the same thread and therefore employ a thread-safe mapping in the form of a \textsf{ConcurrentHashMap}. This, however, raises the challenge of providing easily computable, stable and unique keys. If we consider our taint tags not to store the taint value of a certain value, but of a certain location, we can compute stable identifiers for fields and use them as keys. For \textsf{static} class fields, identifying the specific class and field is sufficient and can be precomputed during compilation. The current implementation injects the computed key as a constant into the \hgraph and provides it as an argument to a field \textit{taint-set} or \textit{taint-get}. For object fields, we do not only need to identify classes but concrete objects, which requires runtime information. In this case, we only inject the field identifier as a constant and provide it together with the field's concrete object to a newly added \textit{TaintLib} function. The returned key is robust to object aliasing, so we do not loose track of objects in e.g. collections. Afterwards, we can use this key in a \textit{taint-set} or \textit{taint-get} for the object field.

It is important to note that our approach to taint tracking depends not only on the entity for which we store taints (i.e., variable locations instead of values), but also on the type of data to which we assign taint values. In our model, we track taints only for primitive types and the taint tag of objects is transitively given by their field's tags. In case of non-primitive fields, the rule applies recursively because eventually all objects can be decomposed to primitives. This design decision is motivated by the fact that tracking all \textit{taint-set} and \textit{taint-get} operations on fields and on all method invocations is more fine-grained than storing taint information at the object level.

\subsection{Further Use Cases}

\paragraph{Dynamic analysis.} Compiler-based solutions are inherently well-suited for white box approaches that require an understanding of the application's internals. One example is the taint tracking \textit{Module} described above that re-instantiates TaintDroid-inspired intra-app taint tracking. Other examples are existing works on commodity systems~\cite{compilerTT,araujo2015compiler} that already utilize compilers for information flow control, which can now be realized on Android as well.

\paragraph{Container solutions.} Modifications of the Android runtime environment have been used in the past (for instance \textit{Divide}\footnote{http://www.divide.com}, now part of Google Android for Work) to establish container solutions that, e.g., encrypt file system I/O of apps or restrict inter-application communication. Using a compiler-based approach such as \name, similar container solutions can be established by replacing the corresponding method invocations (e.g., calls to Java's I/O classes) with calls to injected security-enhanced versions of the same.

\paragraph{Code replacement and compile-time patching.} Google has recently started separating security-critical libraries, such as the notorious WebKit, from application packages into stand-alone apps that are called by apps on-demand. This allows Google to maintain those libraries on an ecosystem-wide scale and roll out security patches more effectively. Since \name is not only able to inject but also to replace or remove code from an app's code base, \name can also be used to apply \textit{compile-time patches} by replacing vulnerable libraries within apps with fixed versions. In an extreme case, this mechanism could allow for removing entire libraries by mocking all their method invocations (e.g., removing ads), or moving code partitions behind a strong security boundary, such as a dedicated process, and reconnect the code through inter-process communications (e.g., as done in the AdSplit~\cite{Shekhar:2012:ASS:2362793.2362821} or AdDroid~\cite{Pearce:2012:APS:2414456.2414498} solutions).

\paragraph{Beyond security: profiling and debugging.} Besides its application in the security domain, using \name to inject tracing, debugging or profiling code allows to gain additional insights into third party applications. A basic example is the method call-tracing we employ in our robustness evaluation in Section \ref{sec:discussion:artist:eval}.

%%% Local Variables: 
%%% mode: latex
%%% TeX-master: "../paper"
%%% End: 

 \section{Discussion}
 \label{sec:discussion}
 This section evaluates \name and its modules in terms of performance, inherent and implementation-specific limitations, and discusses ideas for future work to overcome those and extend our system. 

\subsection{\bname}
\label{sec:discussion:artist}
We first evaluate and discuss general limitations of the \fullname.

\subsubsection{Robustness}
\label{sec:discussion:artist:eval}
We briefly evaluate \name in terms of its robustness by applying an instrumentation routine to a subset of the top apps from different Google play store categories and observing their execution. For the experiment, \name injects into each method of a target application calls to a carefully crafted tracing method that is merged into the app's code. The tracing method uses stack inspection to determine its caller and prints the corresponding method name to the log. All tests are conducted on a real device (rooted Nexus 5 running Android 6 factory image). Out of 85 non-multidex apps (see Section \ref{sec:discussion:artist:impllimit}), 83 apps  were successfully instrumented  and remained stable when tested (97.64\% success rate), indicating the robustness of \name's instrumentation capabilities. 

Because runtime overhead for instrumented apps largely depends on the concrete code that is injected, we refer to the concrete benchmarks of the \textit{Modules} in Sections \ref{sec:discussion:permmodule:eval} and \ref{sec:discussion:taintmodule:eval} for a performance evaluation. 

\subsubsection{Conceptual Limitations}
\label{sec:discussion:artist:conceptlimit}

\paragraph{Native code support.} \opt operates by design on \dex input only. Bundled native libraries (i.e., C/C++) that are connected via JNI are never transformed into \opt's IR and therefore neither instrumented nor inspected by our prototype. Native code components are a limitation of the attacker model of not only our concept but indeed an open challenge for most of the solutions by Android security research, e.g. code analysis as well as IRM solutions in particular.% A potential solution could be the combination of \name with future binary code analysis and instrumentation tools.

\paragraph{Potential fallback to \dex.} The \oat files produced by \dexoat still contain the original \dex byte code of the app to allow fallback to interpretation mode. Naturally, fallback to interpretation would render our instrumentation of the compiled \dex byte code futile. This fallback is currently limited to app debugging, however, no guarantees exist that such a fallback cannot be triggered maliciously. Similarly, dynamically loaded \dex code~\cite{kruegel:ndss14:dynloading,Grace2012} (e.g., via the \hformat{DexClassLoader}) is by default compiled to native bytecode, but no guarantee can be given that dynamically loaded code cannot fall back to interpretation.

\subsubsection{Implementation Limitations}
\label{sec:discussion:artist:impllimit}

\paragraph{Permanence of instrumentation.} Instrumentation of an app's \oat file might be reverted through an application update or a firmware update where apps are re-compiled. Thus, there exists a window of opportunity for an attacker to start an uninstrumented app after a system or app update. Apps, however, cannot be started programmatically after install/update until the user has started the app manually and both scenarios can be detected by \app via system notifications (i.e., broadcasts). Assuming that the system notifies the \app fast enough in order to re-instrument the updated app before the user manually starts the app, the window of opportunity in which an uninstrumented app is started can be closed.

\paragraph{Deployment strategy.} In order to create a pure application layer solution, our prototype currently relies on the na{\"i}ve approach of requesting elevated privileges to replace the installed app \textsf{oat} file with the instrumented version. We can eliminate this requirement by integrating \name with an application layer only sandboxing solution that provides file system virtualization, such as \textit{Boxify}~\cite{backes2015boxify} or \textit{NJAS}~\cite{bianchi2015njas}, or by resetting the execution environment and replacing loaded libraries using reference hijacking \cite{RefHijacking}. Both approaches enable the manipulation of file paths from the original to the instrumented \textsf{oat} file at application startup time.

\paragraph{Multidex support.} Currently our \name prototype does not support multidex\footnote{\url{http://developer.android.com/tools/building/multidex.html}} apps. Enabling support for such apps would not only improve coverage of our prototype for more complex apps, but also simplify the merging of additional code by shifting the merge process from the pre-processing step in \app into the \compiler by simply providing the additional \dex files as input to the compilation process.

\subsection{Dynamic Permission Module}
\label{sec:discussion:permmodule}
We briefly evaluate the performance impact of our dynamic permission module and discuss limitations. 

\subsubsection{Evaluation}
\label{sec:discussion:permmodule:eval}
The additional security checks inlined by our \textit{Module} are only inserted before permission-protected SDK method calls, so we cannot rely on benchmark apps, because they rarely trigger the added functionality. Therefore, we evaluate the performance impact of our permission checking code using custom microbenchmarks. Table \ref{table:PermMicrobenchmarks} depicts the results of our measurements for calls that are protected by 3 distinct permissions. The overhead encountered in the microbangemarks ranges between 1.18\% and 30.65\%, thus showing the feasibility of our prototype.

\begin{table*}[t]
%\resizebox{\columnwidth}{!}{%
\small
\centering
\begin{tabular}{ |l|c|c|c|c| } 
\hline \multicolumn{5}{ |c| }{Microbenchmarks} \\
\hline 
Tested Method & Permission & Baseline & Instrumented & Penalty\\
\hline
WifiManager.getConfiguredNetworks() & ACCESS\_WIFI\_STATE & 0.681 ms & 0.742 ms & 8.89\%\\
\hline 
WifiManager.isWifiEnabled() & ACCESS\_WIFI\_STATE & 0.071 ms & 0.072 ms & 1.18\%\\
\hline
WifiManager.getScanResults() & ACCESS\_COARSE\_LOCATION & 0.452 ms & 0.591 ms & 30.65\%\\
\hline
BluetoothAdapter.startDiscovery() & BLUETOOTH\_ADMIN & 0.910 ms & 0.940 ms & 3.32\%\\

\hline 
%Sum & 61041 & 46649 & \\
%\hline

\end{tabular}
%}
\vspace{+0.2cm}
\caption{Microbenchmarks averaged over 60.000 runs. The \textit{baseline} benchmarks measure the pure execution time of the permission-protected call while the \textit{instrumented} benchmarks measure the protected call and the additional permission check.}
\label{table:PermMicrobenchmarks}
\end{table*}

\subsubsection{Limitations}
\label{sec:discussion:permmodule:limit}

\paragraph{Restriction to synchronous calls.} In order to demonstrate the straightforward implementation of an \name \textit{Module}, we opted for a simple instrumentation strategy that only covers synchronous permission-protected method calls. In result, the current prototype does not support callbacks or asynchronicity and its implementation should therefore be considered a proof-of-concept only. 

\paragraph{Best effort permission map.} In order to direct \name to the instrumentation targets, i.e., the application's permission-protected method calls, we utilize a map of methods calls to the permissions enforced by those calls. While the PScout project \cite{au2012pscout} provides exact API method to permission mappings up to Android version 5.1.1, \name operates on Marshmallow that requires mappings for Android 6. Consequently, the permission map utilized by our \textit{Module} is a hand picked subset of methods from the v5.1.1 PScout map that did not change for Marshmallow.

\subsection{Taint Tracking Module}
\label{sec:discussion:taintmodule}

We evaluate our taint tracking \textit{Module} in terms of feasibility, performance, and the limitations of its current prototypical implementation.

\subsubsection{Evaluation}
\label{sec:discussion:taintmodule:eval}

\paragraph{Runtime overhead.} 
We leverage an Android microbenchmark application to evaluate the performance of our prototype. Since our taint-instrumentation only affects the performance of Java code, we specifically chose the \textit{Passmark} benchmark, which does not contain native libraries and implements all benchmarks in Java. Table~\ref{table:PerfBench} compares the results of the baseline benchmark with a non-instrumented \textit{Passmark} app to those of an instrumented and taint-aware version. The results show an overhead ranging between 7.74\% and 30.73\%, which is within an acceptable range for a taint tracking approach that is not fully tuned for performance. This result is also roughly comparable to microbenchmark results of TaintDroid's~\cite{EnGiBy_10:Taindroid} interpreter-based approach. However, as stated in \cite{managedTT}, microbenchmarks are not very representative in user-driven scenarios such as Android apps, so we take this result with a grain of salt. 

Overall performance can be enhanced by introducing custom optimizations specifically tailored towards improving taint tracking code. One approach would be to eliminate \textit{taint-sets} and \textit{taint-gets} that are based on stack operations and cancel each other out, e.g., alternating \textit{pushs} and \textit{pops} of the same tag as seen for methods that return the return value of another method call. Moreover, the analysis phase allows to abstain from instrumenting apps that do not contain any global taint sinks in order to not impact performance at all in this case.

\begin{table}
%\resizebox{\columnwidth}{!}{%
\small
\centering
\begin{tabular}{ |l|c|c|c| } 
\hline \multicolumn{4}{ |c| }{Passmark} \\
\hline 
Test & Baseline & Taint-Aware & Penalty in \%\\
\hline
CPU & 32521 & 22526 & 30.73\%\\
\hline
Disk & 24893 & 20777 & 16.53\%\\
\hline
Memory & 3627 & 3346 & 7.74\%\\
\hline 
%Sum & 61041 & 46649 & \\
%\hline

\end{tabular}
%}
\vspace{+0.2cm}
\caption{\textsf{Passmark} results averaged over 5 runs.}
\label{table:PerfBench}
\end{table}

\paragraph{Functional Evaluation.} We conducted this case study to research whether intra-application taint tracking can be achieved with an compiler-based instrumentation framework such as \name, so our functional evaluation focuses on detecting different kinds of data leaks in apps. However, to the best of our knowledge, there is no standardized test suite specifically tailored towards evaluating dynamic taint tracking systems for Android apps, and testing real applications is not feasible because they lack the required ground truth. In order to overcome this unsatisfactory situation, we decided to exploit an open source suite called \textit{DroidBench}~\cite{Arzt:2014:FPC:2594291.2594299,droidbenchwebsite} that was initially created to benchmark static taint tracking systems. Even though this does not immediately apply to a dynamic system such as ours, we can still leverage the fact that it provides us with an assortment of applications with different but well-defined leakage behavior. Table~\ref{table:DroidBenchShort} summarizes our \textit{Module's} results for those tests and categories within scope. Tests for implicit flows, inter-component communication, and reflection are omitted because they currently exceed the scope of our proof-of-concept taint tracking. As we are \textit{abusing} the benchmark suite, we need to be careful which conclusions we draw from the test results. The first insight we gain, however, is that our case study succeeded in showing that intra-app taint tracking can be implemented as a pure application layer solution using compiler-driven instrumentation. The second insight we derive is that, as indicated by lower results such as those for the \textit{Android Specifics} category, our proof-of-concept does not yet catch up with previous works such as TaintDroid. Nonetheless, our work not only shows the feasibility of the approach but also lays the foundation for creating a full-fledged taint tracking system for Android versions above Marshmallow that utilizes compiler-based instrumentation and does not require modification of the operating system.

\begin{table}
%\resizebox{\columnwidth}{!}{%
\small
\centering
\begin{tabular}{ |l|c|c| } 
\hline \multicolumn{3}{ |c| }{DroidBench} \\
\hline 
Category & Successful Tests & Ratio \\
\hline
Callbacks & 14/15 & 93\%\\
\hline
Lifecycle & 13/14 & 92.9\%\\
\hline
General Java & 14/20 & 70\%\\
\hline
Aliasing & 1/1 & 100\%\\
\hline
Android Specifics & 5/9 & 55.6\%\\
\hline
Field \& Object Sensitivity & 7/7 & 100\%\\
\hline \hline
Overall & 54/66 & 81.8\%\\
\hline

\end{tabular}
%}
\vspace{+0.2cm}
\caption{Results for the DroidBench taint tracking evaluation. Broken tests and categories not applicable to our system are omitted.}
\label{table:DroidBenchShort}
\end{table}

\newpage

\subsubsection{Limitations}
\label{sec:discussion:taintmodule:limit}

\paragraph{No tracking of implicit flows.} Like TaintDroid~\cite{EnGiBy_10:Taindroid}, our system currently does not track implicit flows (i.e., data leakage using control flow dependencies) and malevolent apps could exfiltrate data in a way that is unnoticeable by our prototype. As the TaintDroid authors discuss, mitigating leakage through control flows would require static analysis and access to the app's source code---both of which TaintDroid could not provide. \name however is already provided with the full app code and it would be highly interesting future work to investigate to which extent the structural program information of the IR and analytical features of the compiler backend (e.g., \opt) can help to remedy the limitations of customary taint tracking solutions on Android.

\paragraph{Taint tracking boundaries.} The compiler is restricted to the app's code base, which introduces imprecision when leaking information through SDK methods, where a \textit{taint-set} at the caller side (developer code) but not the \textit{taint-get} at the callee side (SDK) can be inlined. In particular, and in contrast to object types, storing primitives or strings in collections or sharing them across threads are corner cases where the taints will not be propagated appropriately. This shortcoming can be solved by using pre-computed control-flow models for framework methods~\cite{cao2015edgeminer} to generate corresponding \textit{taint-set} and \textit{taint-get} pairs that model the transition of data through the framework. A preferable technical solution in the future, which removes the potential over-approximations of SDK internal states in control-flow models~\cite{cao2015edgeminer} and which could be of interest beyond taint tracking, is the instrumentation of the \textit{core image}. The core image is a pre-compiled \oat file of the framework classes that is pre-loaded into every application process via Zygote. Since the core image is created with \dexoat during the device startup once after each system update, it can be instrumented using a \compiler as in \name. However, in either case and as in the original work~\cite{EnGiBy_10:Taindroid}, data that already left the phone (e.g., through a network socket) cannot be tracked.

\newpage

\paragraph{Inter-application communication.} Our prototype is currently limited to \textit{intra}-application tracking and lacks support for \textit{inter}-application tracking, for instance, through the file system or Binder IPC. This opens the possibility of confused deputy~\cite{PoWaMoHaCh_11:PermRedeleg,DaDmSa_10:PrivEscalation} or collusion attacks~\cite{ScZhZhInKaWa_11:Soundcomber,BuDaDm_12:TowardsT} to exfiltrate data. Assuming that all installed apps are instrumented, a fix to this problem would be the instrumentation of the I/O method calls in order to write out taints together with the data (e.g., into a file or Binder Parcel) and restore the taints at receiver side. When abandoning the requirement for a pure application-layer solution, our system could also be complemented with the original TaintDroid file system and IPC infrastructure, which is unaffected by the loss of DVM, in order to track taints across applications.

%%% Local Variables: 
%%% mode: latex
%%% TeX-master: "../paper"
%%% End: 

\section{Conclusion}
\label{conclusion}
In this paper, we presented \name, a \dexoat compiler-based instrumentation solution for Android applications that operates at application-layer only. In order to be able to design and implement \name, we first and foremost had to thoroughly study the yet uncharted internals of the new compiler suite and in particular of its \opt backend. A deeper understanding of this compiler suite and the new ART runtime is insofar of interest for the security community, since ART and \dexoat replaced the interpreter-based runtime (DVM) of Android versions prior to Lollipop (i.e., version 5) and hence also voided applicability of any security solution that relies on interpreter instrumentation (e.g., TaintDroid~\cite{EnGiBy_10:Taindroid} and its derivatives~\cite{HoHaJuScWe_2011:AppFence, droidbox}). We study feasibility of our approach through implementing two distinct use cases. Furthermore our case study highlights the capability of a compiler-based instrumentation framework to re-instantiate basic taint tracking for Android apps on the application layer. In general, our results provide compelling arguments, such as higher robustness and better integration, for preferring compiler-based instrumentation over alternative bytecode or binary rewriting approaches.

%%% Local Variables: 
%%% mode: latex
%%% TeX-master: "../paper"
%%% End: 

{

\bibliographystyle{abbrv}
\bibliography{references}
}

\end{document}